# The identification of lead ingots from the Roman mines of Cartagena (Murcia, Spain): the role of lead isotope analysis

By

P.R.Trincherini[1], C. Domergue[2], I. Manteca[3], A. Nesta[4], P. Quarati[5]

[1] *P.R.Trincherini, LIMS (Laboratory for Isotopic Mass Spectrometry), Piazza Martiri di Trarego 8, 28921 Verbania (VB), Italy*
[2] *C.Domergue, UMR 56 08 CNRS, Maison de la recherche, Université de Toulouse-le Mirail, 5 allée Antonio-Machado, 31058 Toulouse Cédex 9, France*
[3] *I.Manteca, Departamento de Ingeniería Minera, Geológica y Cartográfica, Universidad Politécnica de Cartagena, Murcia, España*
[4] *A.Nesta,Dipartimento di Fisica ,Politecnico di Torino, Italy*
[5] *P.Quarati, Dipartimento di Fisica , Politecnico di Torino and INFN , Sezione di Cagliari , Italy*

***Abstract:* Lead isotope analysis is applied to Roman lead ingots to identify and recognise the importance of those from the mines of Cartagena, and to define the cluster characteristics of the Cartagena ore district on an isotopic map.**

## *Introduction*

The metallic deposits of the Iberian peninsula were exploited very early on and drew the attention of many Mediterranean peoples. As early as the Bronze Age, the Sierra Morena copper deposits were exploited. The silver and copper ores of the south-west (Riotinto, Tharsis, and so on) were worked by the Tartessians for the Phoenicians ($8^{th}$-$5^{th}$ c. B.C.) while the lead and silver mines of the Sierra Morena and those near *Carthago Noua* (Cartagena) were worked by the Iberians for the benefits of the Greeks and Carthaginians ($5^{th}$-$3^{rd}$ c. B.C.). Thus when, following the Second Punic war (218-210 B.C.), the Romans begin to conquer Spain, the wealth in metal of the peninsula was well known. They took hold of *Cathago Noua* and the nearby lead and silver mines and then extended the conquered territory by stages, waging war against the Lusitanians and the Celtiberians (153-133 B.C.), capturing Numantia (133 B.C.), fighting Sertorius (83-72 B.C.), and finally under Augustus campaigning against the Asturians and Cantabrians (25-19 B.C.), all the time undertaking to exploit the country’s resources, especially mineral. In the eyes of Rome, *Hispania* was nothing less than an Eldorado, just as America would be for Spain in the $16^{th}$ and $17^{th}$ c.

Of the metal-bearing deposits in Spain, the silver bearing galena mines of the Sierra Morena in N Andalusia and those in the vicinity of the town of *Carthago Noua* in SE Spain provided the conquerors with large amount of silver and lead. Ancient authors, especially Polybius (*History*, 10.2.20; 10.8; 10.38; 34.9.8), Appian (*Iber*.19 and 23), and Livy (23.45.9; 24.41.7; 26.42-27.18; 28.3.2-3; 32. 28.11), spoke of these mines at the time of the conquest, mentioning their extraordinary wealth (Polybius *ap*. Strabo 3.2.10; Livy 34.21.7); two other authors, Diodorus of Sicily, (5.35-38), and Strabo (3.2.3, 3.2.8, 3.2.10-11, 3.4.6) often took their information from

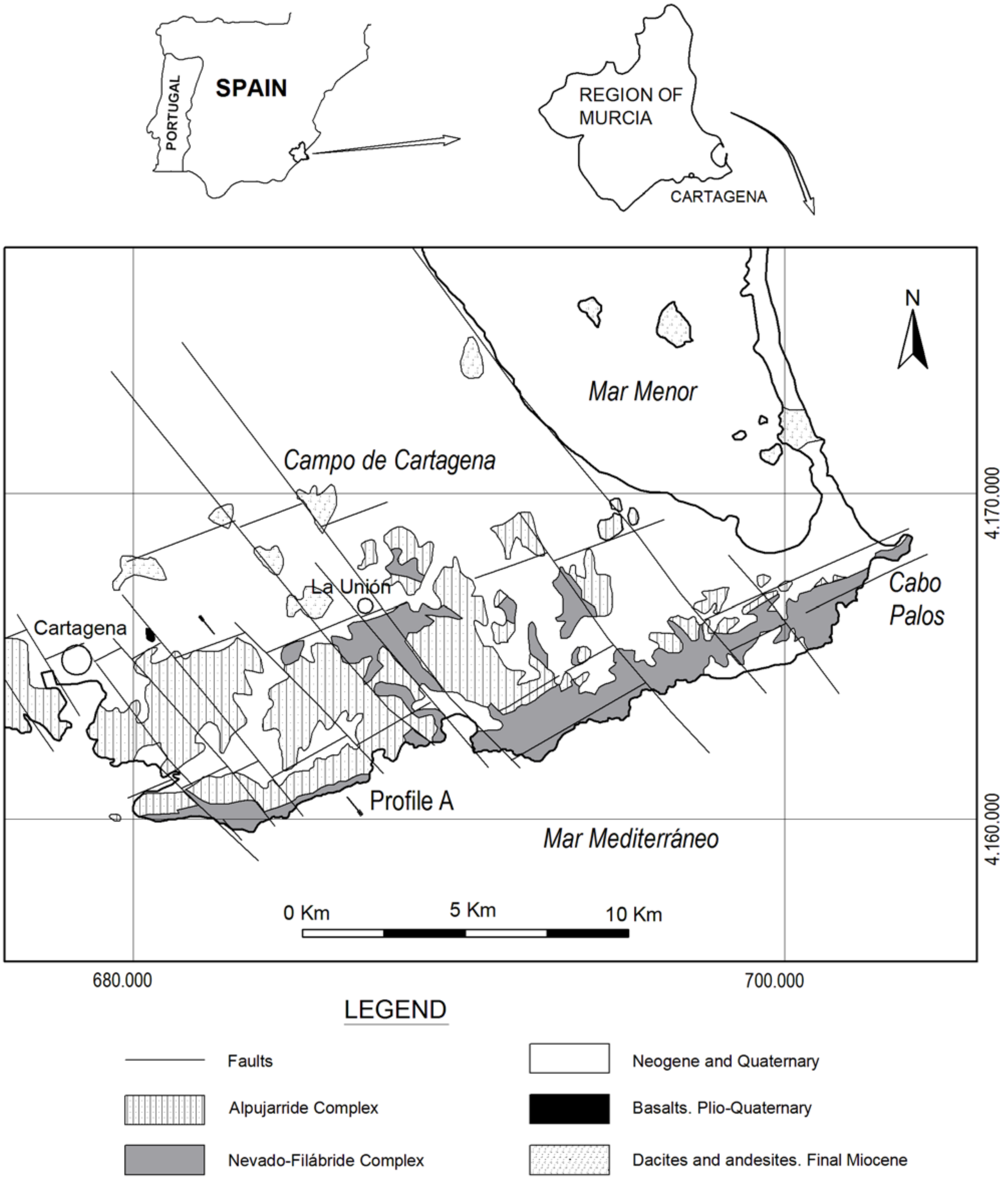

Fig. 1a. Simplified geological map of the Cartagena area, with location of the transverse profile of fig. 1b.

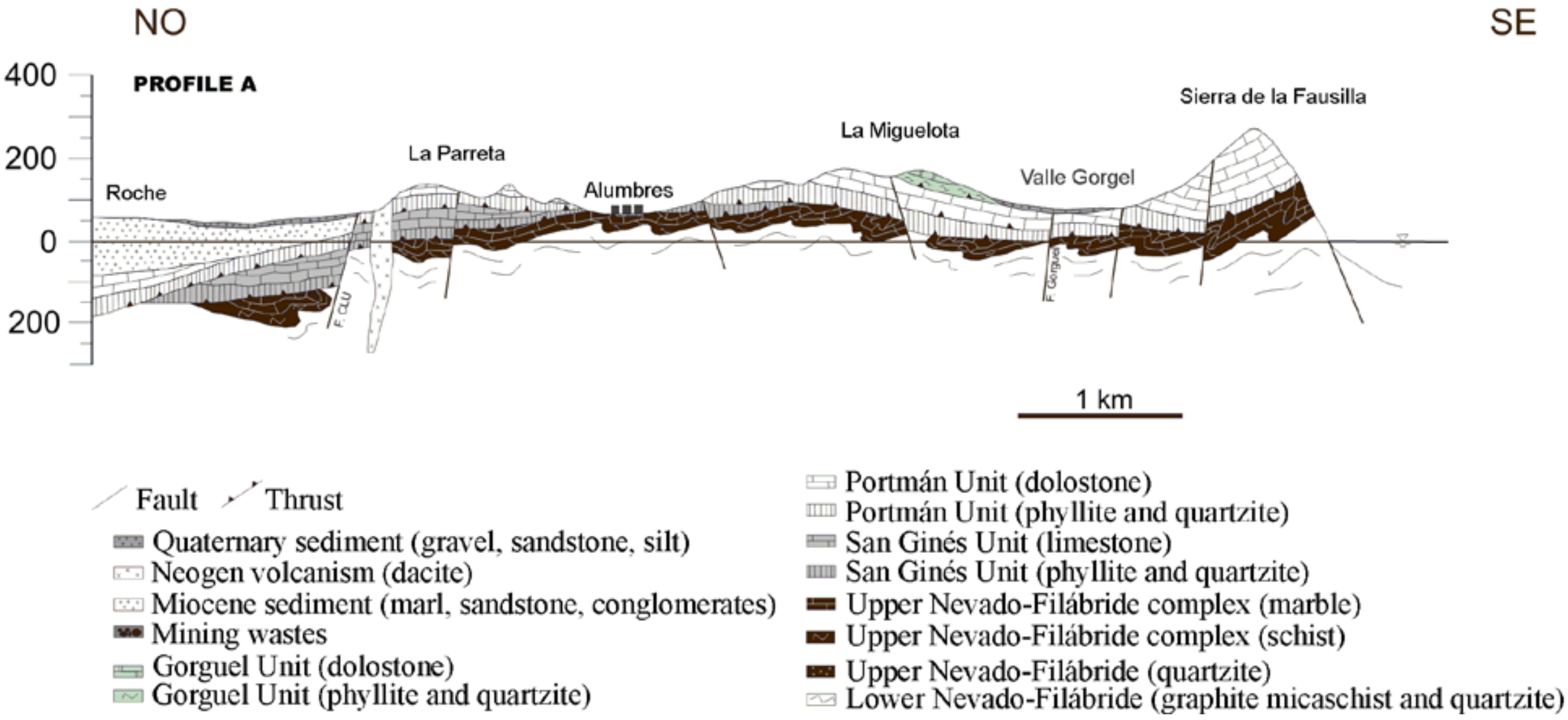

Fig. 1b. Transverse geological profile of the Cartagena mountain range, showing the relation between the different geological unities.

Posidonius, who travelled in these regions at the start of the 1st c. B.C. Archaeology shows that the period of greatest exploitation was between the end of the 2nd c. B.C. and the middle of the 1st c. A.D. (Domergue 1990, p. 179-224). This period happened to be one in which Rome needed both silver, for minting large quantities of *denarii*, its standard currency, and lead, to satisfy growing requirements in the economy, two sectors of which — ship-building and laying out towns — were particularly high consumers of lead. The growth of trade required increasing amounts of lead to fit out ships for the high seas, particularly for lining hulls with lead, as was the norm at the end of the Republic and beginning of the Empire, while at the same time Rome and the major towns of the Empire were in expansion mode, building grandiose public edifices (particularly baths) and private mansions for which running water was a necessity. If stone-built aqueducts brought water to the edges of a city, the pipes which distributed the water within the city were made of lead. It was Spain which could meet the comprehensive need for the lead by Rome and her empire[1]. Plying along the routes of *Mare Nostrum*, ships loaded with the valuable metal brought it from Spain mainly to Italy. Some of the ships sank, and their wrecks mark out the sea routes. Exploration has brought to light cargoes of lead ingots, some of them amounting to tens of tonnes: the wreck off the island of Mal di Ventre, on the W coast of Sardinia, dating from the middle of the 1st century B.C.[2], contained about 1,500 lead ingots from *Noua Carthago*, some45 tonnes (Salvi, 1992).

These ingots bear usually the producers' mould-marks, epigraphy study of which may allow one to discover the origin of the ingots, although many are not amenable to this approach. The analysis of lead isotopes and their relationships is a reliable and fruitful archaeometric procedure to be used in conjunction with study of the epigraphy and the typology of the , and consideration of the archaeological context (in the case of wrecks, the composition of the rest of the cargo and the fittings on board: see e.g., Bernard and Domergue 1991; Domergue 1994). This yelds a more satisfactory approach to the problem of the origin of those metal semi-products, the ingots, desined for trade (Domergue 2008, p. 73-78). The method is described below.

[1] At that time the lead deposits of britain, germany and the Balkans were not yet under Rome's control.
[2] The dates of the wreck are given according to Parker 1992.

Polybius (34.9.8, quoted by Strabo 3.2.10) makes explicit mention of the mines of *Noua Carthago*; he emphasizes the extent of the deposits, the large number of miners working in them, and the amount of profit the State obtained from them. This leads us to think that, among all the Hispanic lead and silver mines, those of *Noua Carthago* played a major rôle in the Roman world at the end of the 2nd and in the 1st c. B.C.

We have had the chance to examine several Roman lead ingots produced in the mines of *Hispania*. By applying the lead isotope method we can better identify those from the mines of *Noua Carthago*, and can gain a better appreciation of their production.

## The Cartagena lead deposits

*Geological summary*

The Sierra de Cartagena is a coastal mountain range located south of the region of Murcia and east of Cartagena, with a length of 23 kilometres and altitudes below 400 m. It contained one of the largest Pb-Zn accumulations in the peninsula (Oen et al, 1975). The range belongs to the Internal Zones of the Betic Cordillera, which emerged together with the Rift Cordillera during the convergence between the African and Iberian plates (late Mesozoic to Cenozoic) (Sanz de Galdeano 1990).
The Sierra de Cartagena shows three distinct major geological complexes superimposed on one another: from bottom to top, the Nevado-Filábride Complex (NFC), the Alpujárride Complex (AC), and the Neogene (N). The NFC comprises strongly-folded metamorphic rocks, with a lower series of graphite micaschists and greyish quartzites of Paleozoic age (LNF), representing the substrate of the Sierra, and an upper formation made up of micaschists, quarzites, marbles and green rocks, of Permo-Triassic age (UNF). The AC is formed of slightly metamorphosed rocks overlying the NFC, and comprises three nappes structural units: from bottom to top, the San Ginés Unit, the Portman and the Gorguel Unit. The latter appears only locally in the area of the Gorguel valley (see the geological profile, fig. 1b). Each of these presents a basal detritic formation (quarzites and phyllites) of Permian-Triassic age and an upper carbonate series (limestones and dolostones) of Middle Triassic age. Intrusive bodies of dolerites are present in the carbonates of the San Ginés Unit (Manteca and Ovejero 1992, Ovejero *et al*. 1976).
The Neogene comprises a marine sedimentary formation (conglomerates, sandstones, siltstones and marls), slightly folded, of upper Miocene age, affected by sub-volcanic rocks (rhyodacites, dacites and andesites) of Pliocene age (7-11 M.Y.) and volcanic formations (alkaline basalts) of Plio-Quaternary age (2.6 M.Y.) (Bellón 1976). To these conjoint must be added recent Quaternary sediments accumulated in valleys and depressions, especially north of the Sierra, in the plain of the Mar Menor (fig.1).

The most distinctive types of mineralization are the *mantos* or strata-bound polymetallic (Fe, Pb, Zn, Ag, Cu, Sn, Mn ) deposits, formed by replacement of Triassic limestones in the Lowest AC or San Ginés Unit (the *primer manto*) and of Triassic marbles in the Upper NF (the *segundo manto*), with great lateral extension and thicknesses of up to 80 metres in the case of the *primer manto* and 25 meters in that of the *segundo manto* (Manteca and Ovejero 1992). There are also numerous major veins. Many lodes have lengths varying between 500 and 1000 metres, with a thickness of *c.*1m. Occasionally thicker veins have been exploited, such as the “Rothschild vein”, up to 10 m thick, or even thicker ones such as the legendary “Cabezo Rajao vein”, as much as 13 m thick and mined since pre-Roman times. In addition there are deposits such as disseminations and stock works; these two types affect Miocene sediments and vulcanites in particular. There are also significant surface gossan-type deposits, formed by oxidation of sulphides.

The main mineral paragenesis in the strata-bound deposits is greenalite-chlorite-magnetite-sulphides-carbonates-silica. The sulphides show a predominance of pyrite, followed by sphalerite, galena, marcasite, and locally pyrrhotite, chalcopyrite, arsenopyrite, tetrahedrite and stannite (Oen *et al.* 1975). The paragenesis in the veins is simpler, with a predominance of galena and sphalerite with carbonates and silica. Galena is always argentiferous and contains some 1.000 ppm of silver in average. In gossan zones ores occur as an oxide-hydroxide-sulphate-carbonate-silica association. The oxides are goethite, haematite, magnetite and manganese oxide; the sulphates include barite, anglesite, jarosite, alunite, and gypsum; the carbonates are siderite, cerusite, and smithsonite; silica appears as quartz, chalcedony and opal, and the clay minerals are vermiculite, metahalloysite, and dickite (Oen *et al*. 1975). The main mineral deposits in the central zone of the Sierra de Cartagena are of epigenetic character. They are of hydrothermal origin related to Pliocene magmatic activity, with a probable chronology between 7 and 11 million years, though they may contain even younger generations of sulphides, through the effect of later remobilizations (Manteca *et al*. 1992). There are also some minor galena deposits in the Portman Unit Triassic dolostones located in peripheral areas of the Sierra, such as the San Julian hill beside the city of Cartagena, which have syngenetic characteristics and might be of Triassic age. This type of galena deposit contains no silver, unlike other deposits in the rest of the Sierra.

*The exploitation of the lead and silver deposits of Cartagena in antiquity*

In the large stratatiform deposits (*mantos*), in which silver-bearing galena predominates, the areas worked to the greatest extent in antiquity were those crossed by more or less rich veins of the silver-bearing galena, which constituted exceptionally rich areas (Domergue 1987, p. 362-380; Orejas, Antolinos 1999; Domergue 2003; Antolinos, Soler 2007). The exploitation of these led to large subterranean workings, called *anchurones* by local miners in the 19$^{th}$-c. miners. The *anchurones* were destroyed by modern open-cast mining in the 20$^{th}$ c., along with most of the ancient shafts and galleries excavated during the underground workings. On the other hand, the ancient mining waste left after the manual sorting of mineral ore by the miners at the pithead was scattered in time over the hillsides near the mine entrances and piled up in the gullies cutting across the mountain. This evidence of ancient mining activity is still there, studded with ancient artefacts (mainly amphoras and earthenware vessels), the study of which has enabled researchers to date the corresponding mining activity to between the 5$^{th}$ and 4$^{th}$ c. and especially in the 2$^{nd}$ - 1$^{st}$ c. B.C. (Domergue 1987, 371-90). The area covered by the débris around the highest point of the Sierra (Sancti Spiritus, elevation 444 m), as well as the remains of ancient galleries which were still visible in 1974 (Domergue 2003, 6-7) on the steps of the modern opencasts (Emilia, San Valentin, Tomasa), suffice to give an idea of the extent of the area worked in antiquity. In addition, settlements and foundries of Roman date are scattered throughout the Sierra. On the plain extending to the north, in the vicinity of La Unión and towards the Mar Menor, the remains of several other Roman metallurgical workshops and agricultural sites have come to light in recent decades during land consolidation and work on orange plantations (Orejas and Antolinos 1999, Orejas and Sánchez-Palencia 2002, 581-89). In short, fieldwork seems to show that in ancient times, and mainly in the Roman period in the 2$^{nd}$ and 1$^{st}$ c. B.C., the intense settlement of this area was linked to the exploitation of the silver-bearing galena of the Sierra de Cartagena.

## Lead ingots from the Cartagena mines: the present state of knowledge based on archaeology and written sources

Some idea of how important mining activity was can also be formed by examining the products of mines, tracing their distribution, making inventories, studying them and extracting the information they contain. Up to now there has been no information about silver ingots coming from the Cartagena mines. On the other hand, quite a number of lead ingots exist from this source, and many of these can be identified by the traditional method (place of discovery, epigraphy). These are the ones we will briefly consider (see Table 1).

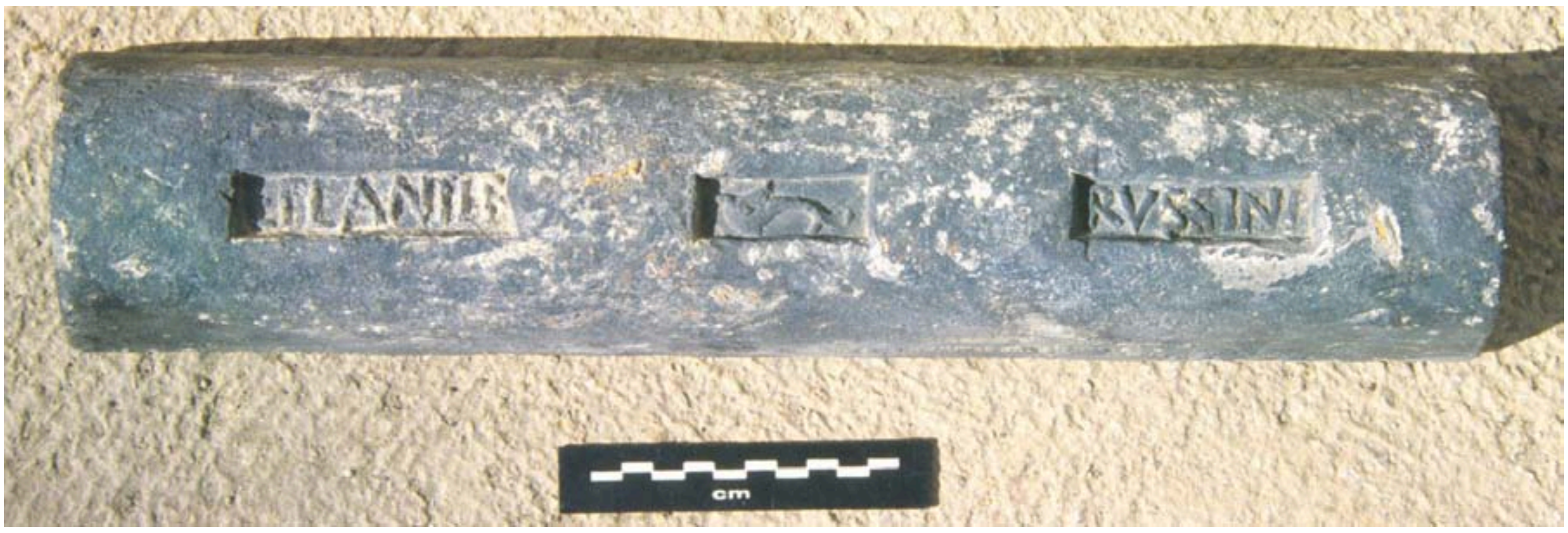

Fig. 2. Typical lead ingot of the silver-lead Cartagena mines from Escombreras 2 wreck (1st c. B.C.). Producer mark: M. PLANI.L.F // delphinus //RVSSINI. Fundación Portuaria, Cartagena (ESC.I/17.17/3/6706 (photo Domergue).

Oblong in shape (fig. 2), they are 10 cm high; they have a rectangular (45 x 10 cm) base. In cross-section they are roughly semi-circular; the short sides are slanted, and they weigh *c.* 32-33 kg, the equivalent of 100 Roman *librae* (32, 740 kg). However, these characteristics (sometimes with variations, for example in weight) are also those of more or less contemporary ingots from the mines of the Sierra Morena, as emerges from the cargoes of the wrecks Cabrera 5 or Sud-Perduto 2 (early 1st c. B.C.) (Colls *et al.* 1986; Bernard, Domergue, 1991; Domergue 2000; Trincherini *et al.* 2001, 397-402). Therefore one must take into account the names of producers (probably the mine owner) recorded in the mould-marks on the ingots[3] (fig.2). By comparing these names with the names backed by epigraphical evidence of *Carthago Noua* (inscriptions and coins, both dated between the 2nd c. B.C. and the 1st c. A.D.), one can assign to the Cartagena mines a series of ingot marks (eg. Domergue 1985; Koch 1994; Abascal, Ramallo 1997). The discovery of other ingots in the mining territory itself (ingots produced by the Roscii, at the caserío de Roches, not far from the Cabezo Rajado), in the port of Cartagena during various exploratory operations (e.g., the dredging of the harbour in 1878, or on the occasion of recent underwater excavations in the outer harbour of Escombreras (Pinedo Reyes, Alonso Campoy 2004) confirms this preliminary information and/or allows the list to be expanded.

At present, the traditional methods of investigation attribute to the mines of *Carthago Noua* ingots produced by individuals called Appuleius, Aquinius, Atellius, Aurunculeius, Cornelius, Dirius, Fiduius, Furius, Gargilius, Iunius, Laetilius, Lucretius, Messius, Nona, Nonius, Planius, Pontilienus, Raius, Roscius, Seius, Turullius, Varius, Vtius. In fact, the variety of stamps with these names appears in a variety of combinations, so that this list of 23 names represents at least 42 different brands and *c.*970 ingots. These figures and the dispersion of the ingots throughout the Roman West provide some idea of the importance of the production and trade in lead from the mines of Cartagena. In addition, other ingots, with poorly characterized features (an unusual form, mould-mark missing or erased, lack of epigraphical comparison), may derive from the Cartagena mines. In order to identify those, we turn to lead isotope measurement.

[3] These producers' habit of marking lead ingots with the names moulded with raised letters in rectangular panels on the upper part of the ingots appears first in Spanish ingots. This habit will be followed by the producers of lead ingots in Britain, Germany and Sardinia.
On the mould-marks of the ingots from *Carthago Noua* the producers sometimes have the *tria nomina*, sometimes only the *duo monima*; filiation is generally indicated, sometimes also the name of the Roman tribe. Some of the producers are freedmen. They are generally individuals, sometimes associates. Figurative symbols (anchor, dolphin, rubber, etc.) may appear in the mould-marks, the variety of which is shown in Tables 1, 3 and 4.

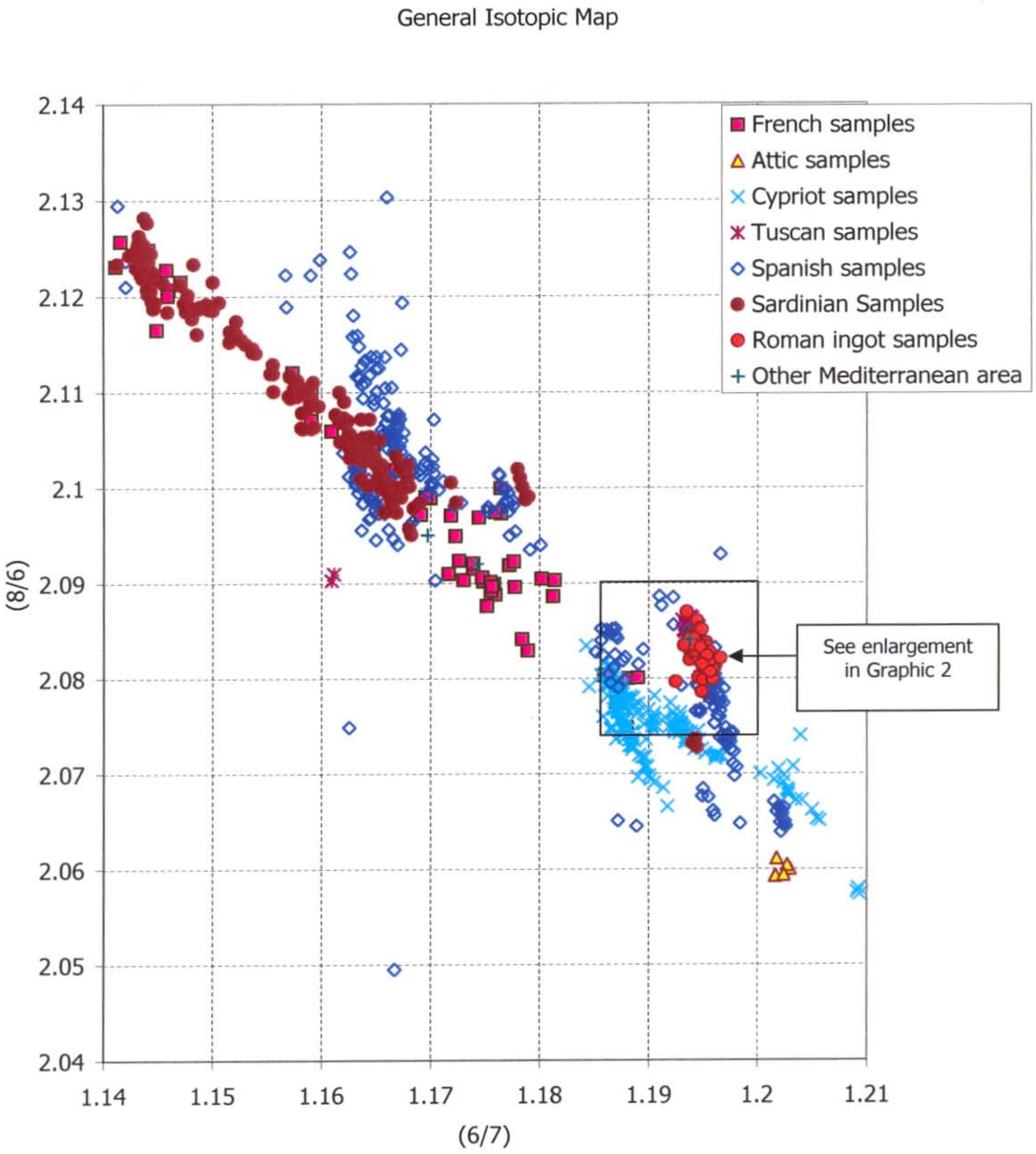

***Graph. 1: General isotopic map***

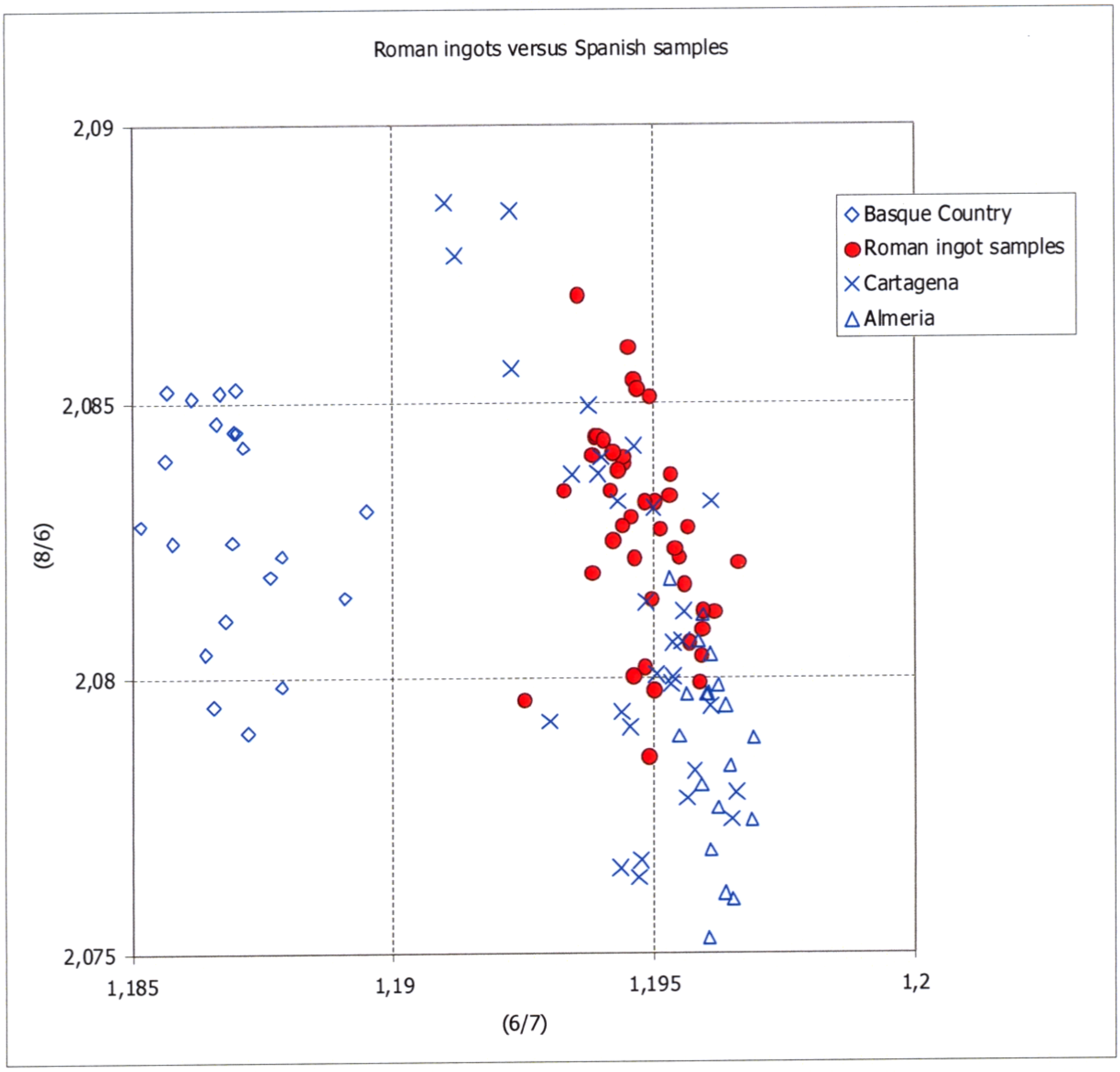

***Graph. 2:* Detail of the general isotopic map. In evidence the Spanish area where, according to the epigraphic studies, the 43 ingots measured are compatible with the mineral existing samples.**

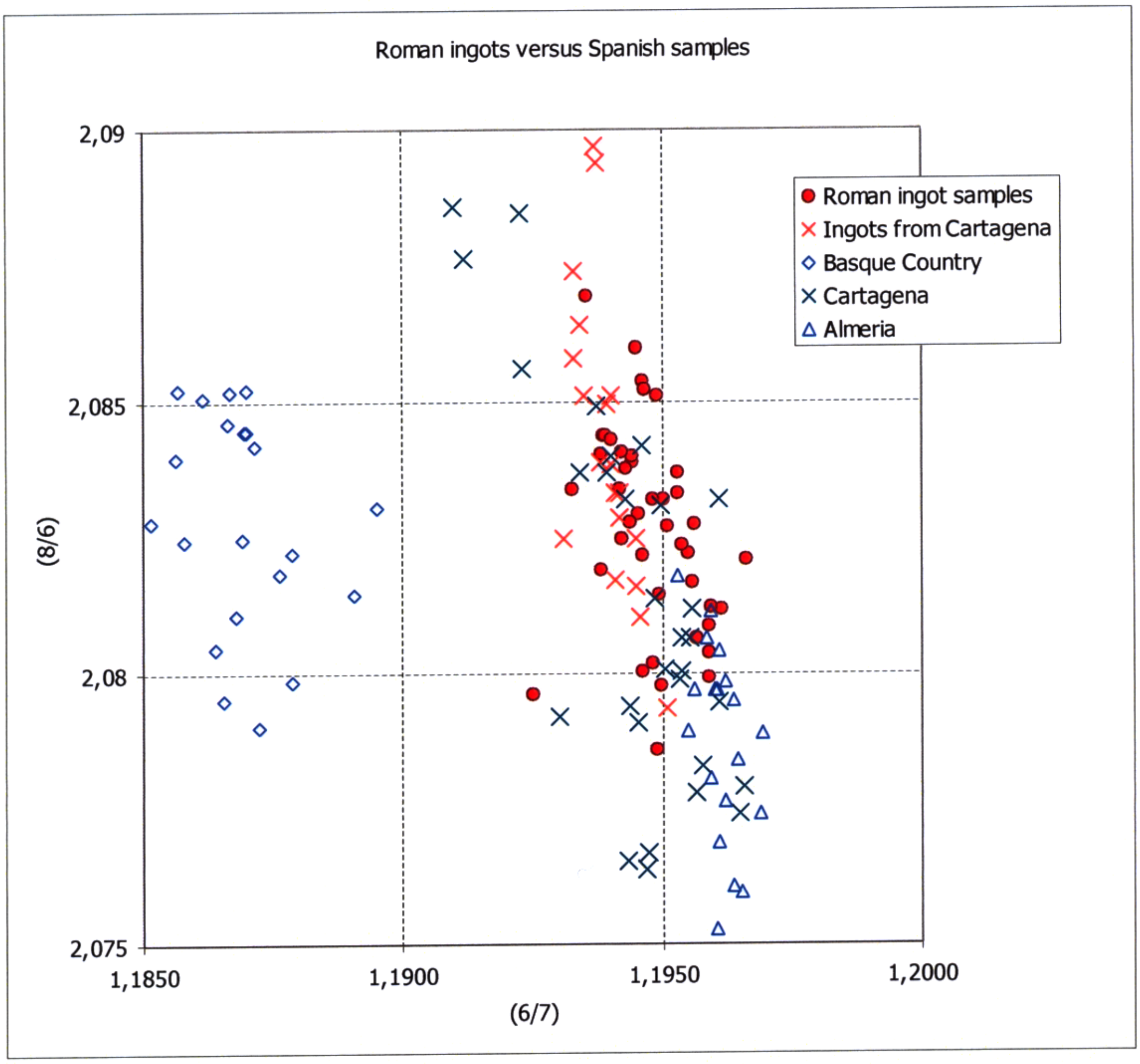

***Graph. 3: 20 ingots from unknown origin have been added to the Cartagena cluster designed in Graph 2. The overlapping confirms the geographical origin from Cartagena ore district.***

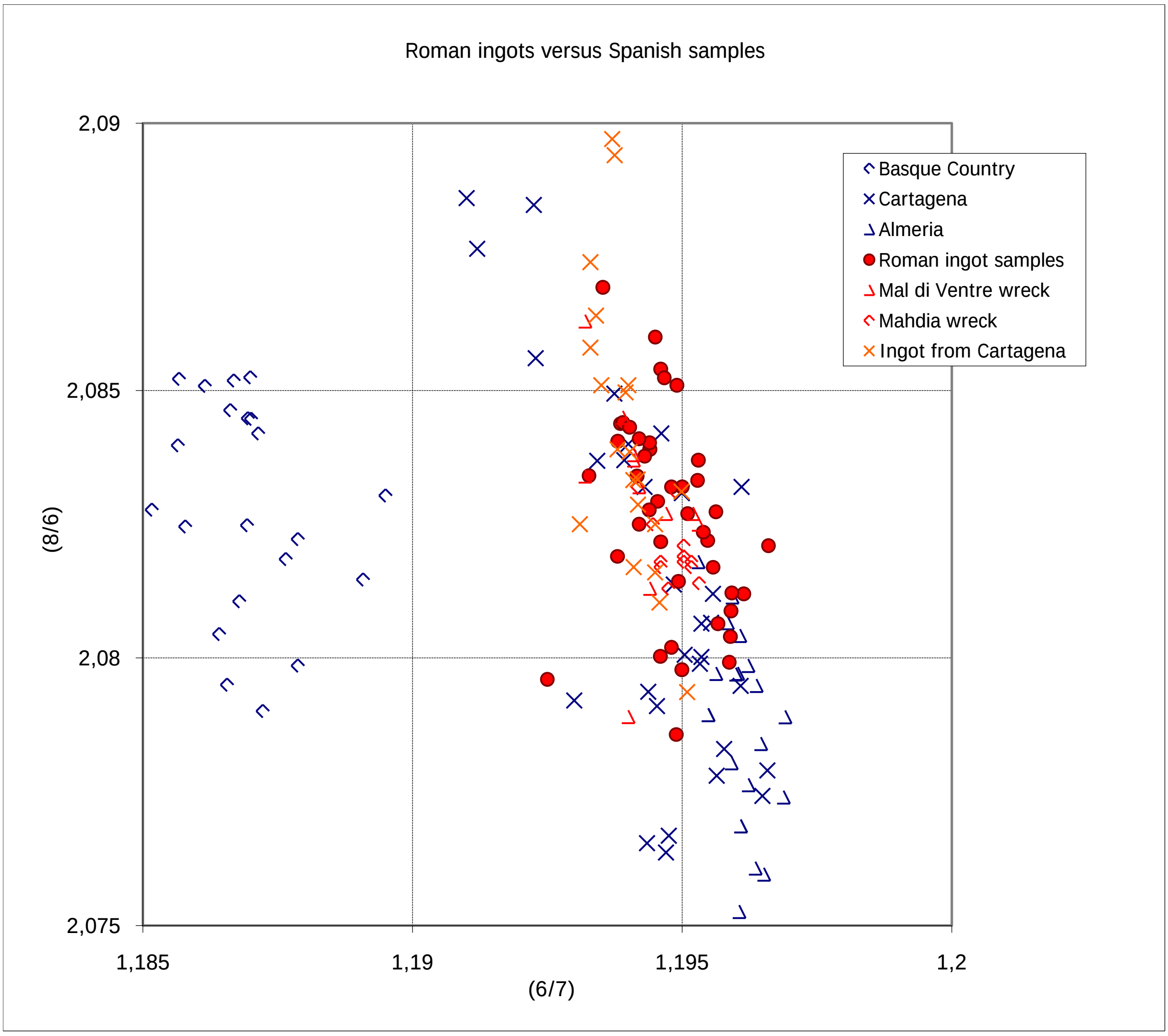

## Instrumentation and isotopic measurements

Lead isotopes were measured on samples of Roman lead ingots from the Cartagena region to explore their relationship. Our main aims being to trace and verify their geographical origin and, if possible, define their route through history with essential support and collaboration from archaeologists. We examined and isotopically measured 43 samples housed in several museums from ingots regarded as originating from Cartagena. Isotopic lead measurements are known to represent a powerful tool in specific archaeometric applications (Gale 1997, Trincherini 2001). The spontaneous decay of two isotopes of natural uranium ($^{238}U$, $^{235}U$) and one of thorium ($^{232}Th$) produces 3 of the existing 4 natural lead isotopes ($^{206}Pb$, $^{207}Pb$, $^{208}Pb$). The geological age of the formation and the fact that these two elements originally had a different content in the minerals forming the earth's crust generate variable isotopic compositions in the minerals from which lead is extracted.

The treatment of the samples under study was performed at the Ispra (I) laboratories of the Joint Research Centre of the European Commission, with the isotopic measurements performed at LIMS (Laboratory for Isotopic Mass Spectrometry), Verbania (I). A Finnigan 262 mass spectrometer was used for the purpose, equipped with a variable multi-collector with extensive optical geometry. Although our samples came directly from ingots, the preliminary chemical treatment of samples like these is of particular importance because all the operations must be performed avoiding any possible lead contamination, whether from the environment or from reagents and materials. Consequently, a cleanroom of a class lower than 100 was used for the pre-treatment of the samples. MilliQ water and ultrapure reagents were redistilled twice, while only ultra-cleaned FEP (FluorEthylPropilene) was used as containment material. To avoid the influence of possible metal impurities 1 mg of lead was sampled and then purified by passage on a microcolumn containing anionic resin, adapting a well-known procedure (Koide and Nakamura 1990).

## Discussion: geology, mines, archaeometry, archaeology and history

Studies in which the measurement of isotopic ratios is undertaken to distinguish such small numerical differences, usually called "isotopic anomalies", demand an extremely high level of analytical precision. More precisely, for this kind of investigation both the statistical variables – *precision* and *accuracy* – necessary to assure good-quality data must be lower than the discrimination value used to separate different known geographical origins. In our case the combination of precision and accuracy, defined as the "final uncertainty", is lower than one digit in the third decimal place, meaning that in general and for all the isotopic ratios measured, the final error is $\leq$ 0,1%. Table 1 gives the results of the measurements obtained on the 43 samples taken from the ingots under examination; the same table reports for each ingot both the identification parameters (mainly epigraphic) and the sites of discovery and preservation.

In a previous study (Domergue *et al*., forthcoming) we used the same equipment and methodology to measure the lead isotope ratios in 20 of the 102 ingots recovered from the cargo of the *Comacchio (I)* wreck. (late 1$^{st}$ c. B.C.) (fig.3). The results gave an isotopic signature corresponding to Cartagena mines for 17 ingots, with the other 3 showing an isotopic signature closer to the Almeria mine district, which is geologically more recent. In the Comacchio study, the question of whether it was possible to characterize two separate geological systems present in the Cartagena ore deposits (one of these being close to the Almeria mines), or whether there was only one system different from that of Almeria, was left unresolved.

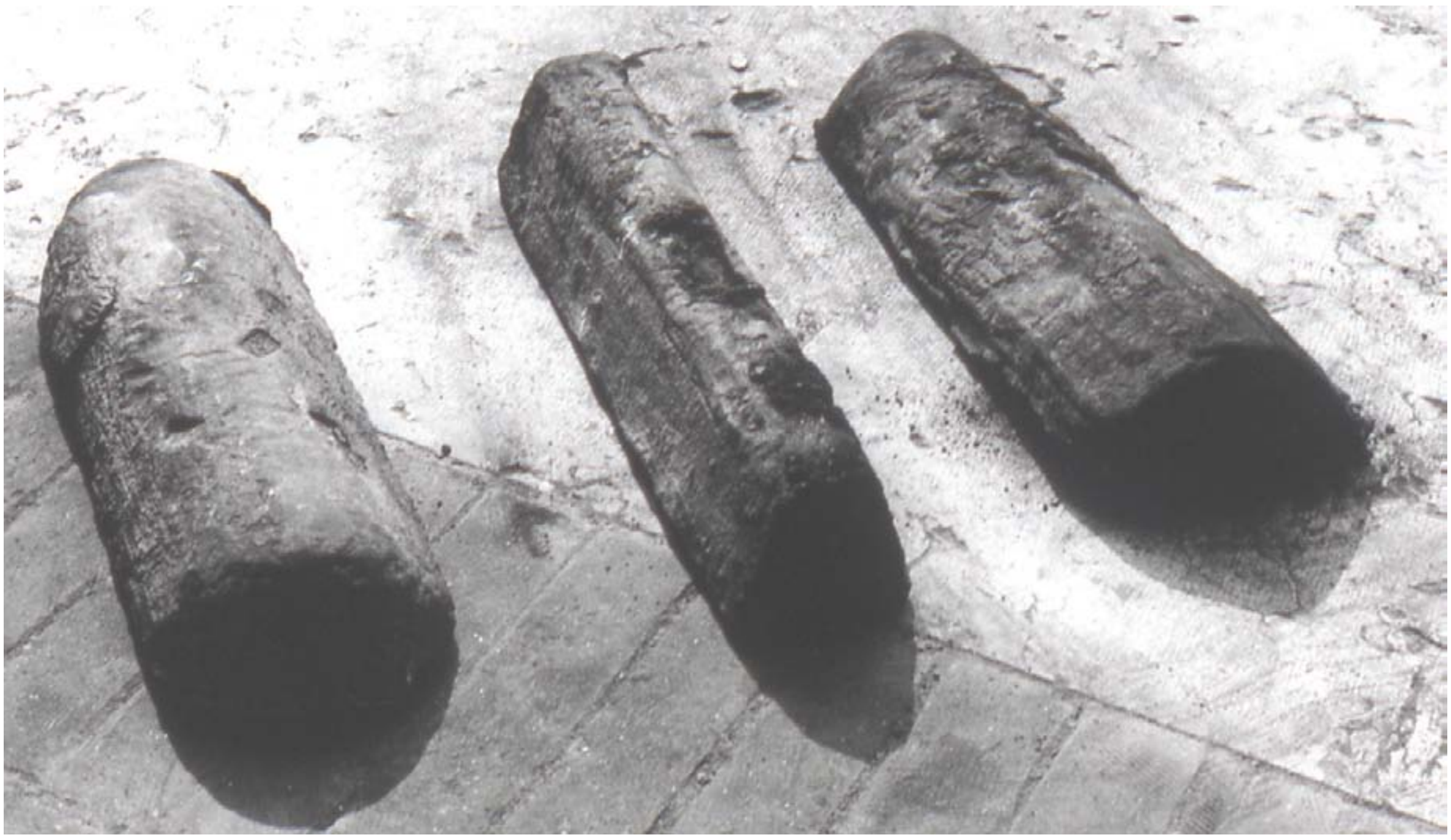

Fig. 3. Comacchio (Ferrara) ingots (age of Augustus). Museum of the Roman Ship, Comacchio (photo C. Domergue).

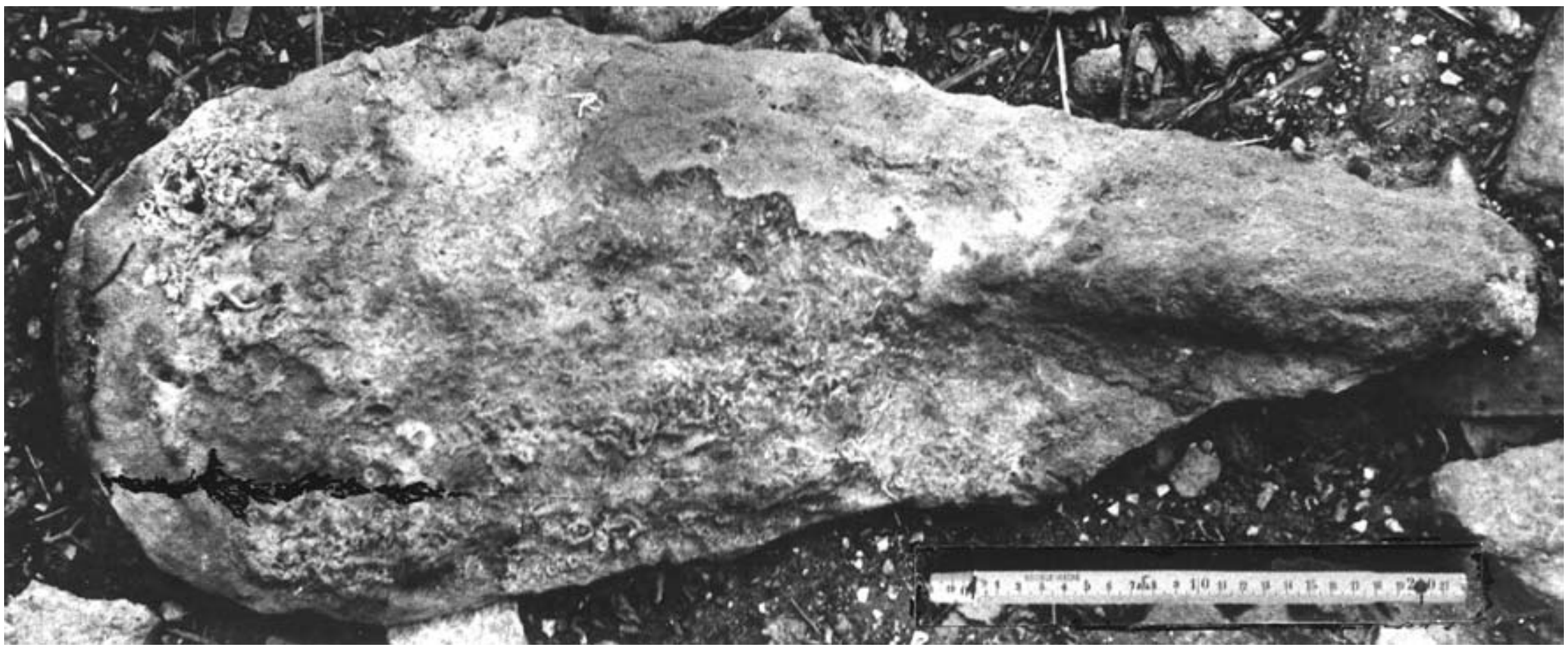

Fig. 4. Lead ingot of the Pinna Nobilis L. Type from *Cabrera 2* (4th-3rd c. B.C.). Private collection (photo C. Domergue).

In the present study we measured 43 ingot samples presumably of Cartagena origin (insofar as they all came from manufacturers known at Cartagena) with the aim of verifying their geographical origin through their isotopic signature and then including the results on the isotopic map of SE Spain. The results are graphically reported on a general isotopic map (Graph.1) and in greater detail in Graph. 2.

Of some 300 mineral and ingot samples recently measured (the results are not yet published), a series of 13 Roman and 7 pre-Roman ingots with no epigraphic information nor any indication as to their geographical origins were found to fall into the Cartagena cluster by virtue of the total consistency of their lead isotope ratios with the amount of data now available. Table 2 (Roman ingots and pre-Roman ingots) reports the isotopic results of tests on these 20 samples, while

| *N°* | N°. sample | Producer | Place of discovery and date of the wreck | Site of preservation | R 206/207 | R 208/206 | R 206/204 | R 207/204 | R 208/204 |
|---|---|---|---|---|---|---|---|---|---|
| *1* | **98/14** | ***delphinus*** **/ / C.AQVINI / /** ***ancora*** | *Escombreras 2* (1st cent. BC) | Museo Arq. Mun.Cartagena | **1,1950** | **2,0832** | **18,719** | **15,665** | **38,995** |
| *2* | **98/35** | **"** | “ | Fundación portuaria Cartagena | **1,1946** | **2,0854** | **18,758** | **15,703** | **39,116** |
| *3* | **98/42** | **M. AQVINI.C.F.** | *Bajo de Dentro* (1st cent. BC) | Museo Naval, Madrid (n°,1064) | **1,1944** | **2,0839** | **18,707** | **15,662** | **38,984** |
| *4* | **00/41** | **"** | “ | “ (n° 1064) | **1,1959** | **2,0804** | **18,725** | **15,658** | **38,959** |
| *5* | **00/42** | **"** | “ | “ (n° 1063) | **1,1961** | **2,0812** | **18,737** | **15,665** | **38,996** |
| *6* | **00/48** | **CN.ATELLI.CN.F.MISERINI** | Carta Romana (Ischia) (no wreck) | Villa Arbusto (Ischia) | **1,1959** | **2,0812** | **18,742** | **15,672** | **39,007** |
| *7* | **98/07** | **L. AVRVNC L.L.AT** | Mines of Huelva (no wreck) | Museo Provincial, Huelva | **1,1938** | **2,0819** | **18,671** | **15,640** | **38,872** |
| *8* | **00/77** | **"** | “ | “ | **1,1939** | **2,0844** | **18,722** | **15,682** | **39,025** |
| *9* | **98/44** | ***lunula*** **/ / M.SEX.CALVI.M.F. / /** ***lunula*** | *Bajo de Dentro* (1st cent. BC) | Museo Naval, Madrid (n°1062) | **1,1953** | **2,0837** | **18,732** | **15,671** | **39,032** |
| *10* | **00/43** | **"** | “ | “ | **1,1959** | **2,0809** | **18,724** | **15,652** | **38,999** |
| *11* | **00/110** | **"** | “ | Museo Nac. Arq. Maritima, Cartagena (n° 50195/Pb 51) | **1,1953** | **2,0833** | **18,742** | **15,680** | **39,046** |
| *12* | **95/38** | **L.CARVLI.L.F.HISPALI MEN** | Crestaig. La Puebla (Mallorca) (no wreck) | Museo Provincial (Mallorca) | **1,1939** | **2,0844** | **18,716** | **15,677** | **39,011** |
| *13* | **98/08** | **L.CARVLI.L.F.HISPALI MEN** | *Madrague de Giens* (70-50 BC) | DFS (Giens, Var) | **1,1942** | **2,0825** | **18,701** | **15,661** | **38,945** |
| *14* | **98/15** | **C.FIDVI.C.F / / S.LVCRETI.S.F** | *Escombreras 2* (1st cent. BC) | Museo Arq. Mun.Cartagena | **1,1948** | **2,0802** | **18,712** | **15,661** | **38,924** |
| *15* | **00/109** | **“** | *Escombreras 2* (?) (1st cent. BC) | Museo Nac. Arq. Maritima, Cartagena (n°88035) | **1,1945** | **2,0829** | **18,720** | **15,672** | **38,994** |
| *16* | **00/02** | **A.P.FVRIEIS.C.P.L.L** | Martigues (Bouches-du-Rhône) (no wreck) | Musée du Vieil Istres, Bouches-du-Rhône !n° D- | **1,1949** | **2,0786** | **18,699** | **15,650** | **38,867** |
| | | | | 2985) | | | | | |
| *17* | **00/19** | **SOC. / / L.GARGILI.T.F. ET.M.LAETILI.ML / / *delphinus*** | *Off Marseillan-Plage (= Parker, Agde J)* (Late 2nd – early 1st cent. BC) | Musée de l'Éphèbe (Agde) | **1,1940** | **2,0843** | **18,727** | **15,684** | **39,035** |
| *18* | **00/107** | **[.].LVCRETI *hedera* / / SOCIETATVS** | *wreck near Cartagena* (no date) | Museo Nac. Arq. Maritima, Cartagena (n° 729/Pb 6) | **1,1946** | **2,0822** | **18,741** | **15,688** | **39,023** |
| *19* | **00/108** | **"** | “ | “ (n° Pb 47/50191 | **1,1935** | **2,0869** | **18,913** | **15,846** | **39,475** |
| *20* | **03/002** | **SOCIETAT / / S.T.LVCRETI** | Monastery of Klingenthal (Switzerland) (no wreck) | Historisches Museum (Switzerland) (1st part : n°1905-6636 a) | **1,1946** | **2,0800** | **18,705** | **15,658** | **38,907** |
| *21* | **03/003** | **"** | “ | “ (2nd part : n° 1935-6636 b) | **1,1950** | **2,0798** | **18,692** | **15,641** | **38,875** |
| *22* | **98/41** | ***delphinus* / / C.MESSI.L.F** | *Bajo de Dentro* (1st cent. BC) | Museo Naval, Madrid (n° 1061) | **1,1944** | **2,0839** | **18,707** | **15,662** | **38,984** |
| *23* | **00/44** | **"** | “ | “(n° 1061) | **1,1955** | **2,0822** | **18,726** | **15,673** | **38,992** |
| *24* | **98/12** | **P.NONAE.P.F.NVC** | Harbour of Cartagena (dredging 1878) | Museo Arq. Munic. .Cartagena | **1,1949** | **2,0814** | **18,764** | **15,703** | **39,056** |
| *25* | **98/13** | **L.PLANI.L.F / / delphinus / / RVSSINI** | *Escombreras 2* (1st cent. BC) | Museo Arq. Munic. Cartagena | **1,1925** | **2,0796** | **18,682** | **15,666** | **38,851** |
| *26* | **98/34** | **"** | Ripatransone (Ascoli-Piceno, Italy) (no wreck) | Museo Civico, Ripatransone (n° 668) | **1,1945** | **2,0860** | **18,753** | **15,699** | **39,118** |
| *27* | **98/45** | **"** | *Escombreras 2* (1st cent. BC) | Museo Naval, Madrid (n° 435) | **1,1959** | **2,0799** | **18,736** | **15,667** | **38,970** |
| *28* | **00/15** | **L.PLANI.L.F.RVSSINI / / *ancora*** | *Off Marseillan-Plage (= Parker, Agde J)* (Late 2nd – early 1st cent. BC) | Musée de l'Éphèbe (Agde) (n° 601) | **1,1944** | **2,0840** | **18,747** | **15,694** | **39,069** |
| *29* | **00/45** | **L.PLANI.L.F / / delphinus / / RVSSINI** | *Escombreras 2* (1st cent. BC) | Museo Naval, Madrid (n° 438) | **1,1956** | **2,0817** | **18,738** | **15,671** | **39,001** |
| *30* | **00/46** | **L.PLANI.L.F / / delphinus / / RVSSINI** | *Escombreras 2* (1st cent. BC) | “ (n° 435) | **1,1957** | **2,0806** | **18,743** | **15,676** | **39,000** |
| *31* | **00/14** | **C.PONTILIENI.S.F. / / *delphinus*** | Baie de la Roquille | Musée de l'Éphèbe | **1,1938** | **2,0841** | **18,712** | **15,674** | **38,999** |
| | | | (Marseillan) | (Agde) (n° 600) | | | | | |
| *32* | **00/17** | **SOC.M.C.PONTILIENORVM.M.F** | *Off Marseillan-Plage (= Parker, Agde J)* (Late 2nd – early 1st cent. BC) | Musée de l'Éphèbe (Agde) (D.1103) | **1,1943** | **2,0838** | **18,715** | **15,671** | **38,999** |
| *33* | **00/18** | **"** | “ | Musée de l'Éphèbe (Agde) (D.1102) | **1,1944** | **2,0828** | **18,713** | **15,667** | **38,976** |
| *34* | **98/11** | **M.RAI.RUFI / / FER** | Harbour of Cartagena (dredging 1878) | Museo Arq. Munic. Cartagena | **1,1933** | **2,0834** | **18,763** | **15,724** | **39,091** |
| *35* | **98/17** | **M.P.ROSCIEIS.M.F.MAIC** | Caserío de Roches, La Unión, Murcia) (no wreck) | Private collection Cartagena) | **1,1942** | **2,0841** | **18,710** | **15,667** | **38,994** |
| *36* | **00/08** | **"** | “ | Louvre (Paris) n° 01.038?10 | **1,1942** | **2,0834** | **18,711** | **15,669** | **38,987** |
| *37* | **00/49** | **"** | “ | Cabinet des Médailles (BN, Paris) n° 2408 | **1,1956** | **2,0827** | **18,728** | **15,664** | **39,006** |
| *38* | **00/111** | **Q.SEI.P.F.MEN.POSTUMI** | *Nido del Cuervo* (Aguilas, Murcia) (1st cent. BC) | Museo Nac. Arq. Maritima, Cartagena (n°50196/Pb 52) | **1,1947** | **2,0852** | **18,758** | **15,702** | **39,116** |
| *39* | **95/39** | **P.TVRVLLI.M.F. / / MAI *delphinus*** | Harbour of Cartagena (before 1896) | Museo Arqueológico Provincial (Murcia) | **1,1966** | **2,0821** | **18,730** | **15,653** | **38,997** |
| *40* | **98/16** | **Q VARI HIBERI** | Harbour of Cartagena | Private collection Cartagena) | **1,1951** | **2,0827** | **18,704** | **15,650** | **38,955** |
| *41* | **98/09** | **C.VTIVSC.F / / *delphinus*** | *Madrague de Giens* (70-50 BC) | DFSM, Hyères (Var) | **1,1948** | **2,0832** | **18,699** | **15,650** | **38,954** |
| *42* | **98/43** | ***delphinus* / / C.VTIVS.C.F. / / *caduceus*** | *Bajo de Dentro* (1st cent. BC) | Museo Naval, Madrid (n° 584) | **1,1949** | **2,0851** | **18,754** | **15,694** | **39,104** |
| *43* | **00/47** | **"** | “ | “ | **1,1954** | **2,0824** | **18,733** | **15,671** | **39,008** |

***Table 1. Lead isotope ratios measured in 43 Roman ingot samples epigraphically identified as manufactured in Cartagena ore district. Name of producer, site of discovery, date of the wrecks and site of preservation are reported. The date of the wrecks is given by the composition of the cargoes (see Parker 1992). Isotopic ratios were measured by LIMS.***

| N° | Sample n° | Rest of mould-mark | Site of discovery and date of the wreck | Site of preservation | $R_{i/j}$ 206/207 | $R_{i/j}$ 208/206 | $R_{i/j}$ 206/204 | $R_{i/j}$ 207/204 | $R_{i/j}$ 208/204 |
|---|---|---|---|---|---|---|---|---|---|
| 1 | 98-29 | […..] // […..] | *Danger d'Algajola (Haute-Corse)* c.. 150-100 BC | Musée d'Ethnographie Corse (Bastia) | 1,1945 | 2,0825 | 18,686 | 15,644 | 38,915 |
| 2 | 98-30 | " | " | " | 1,1945 | 2,0816 | 18,681 | 15,639 | 38,887 |
| 3 | 98-31 | " | " | " | 1,1935 | 2,0851 | 18,728 | 15,691 | 39,051 |
| 4 | 98-32 | " | *Îlot des Moines (Corse du Sud)* c. 1st cent. BC (?) | " | 1,1938 | 2,0839 | 18,716 | 15,678 | 39,003 |
| 5 | 98-33 | " | " | " | 1,1941 | 2,0817 | 18,679 | 15,643 | 38,885 |
| 6 | 98-10 | " | " | " | 1,1933 | 2,0874 | 18,753 | 15,715 | 39,145 |
| 7 | 98-22 | […] // […..] // […] | Off Grande Sanguinaire (Corse du Sud) No date | DFSM Ajaccio (Corse) | 1,1933 | 2,0858 | 18,727 | 15,693 | 39,061 |
| 8 | 98-23 | | Golfe de Ventilegne (Corse du Sud) No date | " | 1,1937 | 2,0897 | 18,786 | 15,737 | 39,257 |
| 9 | 99-24 | […..]I.L.F. [….] | *Madrague de Giens* (70-50 BC) | DRASSM, Marseille, n° 1185 | 1,1950 | 2,0831 | 18,775 | 15,694 | 39,067 |
| 10 | 00-09 | No mould-mark | *Baie de l'Amitié (Marseillan, Hérault)* c. 50-150 AD | Musée de l'Éphèbe, Cap-d'Agde. n° 562 | 1,1941 | 2,0833 | 18,714 | 15,673 | 38,991 |
| 11 | 00-10 | " | " | Musée de l'Éphèbe, Cap d'Agde. n° 571 | 1,1942 | 2,0833 | 18,718 | 15,673 | 39,000 |
| 12 | 00-11 | " | " | Musée de l'Éphèbe, Cap-d'Agde. n° 581 | 1,1942 | 2,0829 | 18,708 | 15,666 | 38,971 |
| 13 | 02-06 | No mould-mark | Riotinto (no wreck) | Museo de Riotinto (Huelva). N° 21 67 | 1,1940 | 2,0839 | 18,728 | 15,685 | 39,027 |

| N° | Sample n° | Site of discovery and date of the wreck | Site of preservation | $R_{i/j}$ 206/207 | $R_{i/j}$ 208/206 | $R_{i/j}$ 206/204 | $R_{i/j}$ 207/204 | $R_{i/j}$ 208/204 |
|---|---|---|---|---|---|---|---|---|
| **14** | **98-50** | ***Cabrera 2* (= Parker 124) *c. 250-225 BC*** | **Monasterio de Lluc (Mallorca)** | **1,1940** | **2,0851** | **18,724** | **15,682** | **39,041** |
| **15** | **98-51** | **"** | **"** | **1,1940** | **2,0850** | **18,720** | **15,681** | **39,033** |
| **16** | **98-52** | **"** | **"** | **1,1931** | **2,0825** | **18,753** | **15,718** | **39,053** |
| **17** | **95-34** | **“** | **Majorca. Private collection** | **1,1934** | **2,0864** | **18,738** | **15,702** | **39,095** |
| **18** | **03-04** | ***Îlot de Brescou, Agde*** **(= Parker *17*) 3rd cent. BC ?** | **Musée de l’Éphèbe, Cap-d’Agde** | **1,1946** | **2,0810** | **18,718** | **15,669** | **38,959** |
| **19** | **03-05** | **“** | **“** | **1,1951** | **2,0794** | **18,691** | **16,640** | **38,866** |
| **20** | **01-01** | **“** | **Musée du Biterrois, Béziers** | **1,1938** | **2,0894** | **18,750** | **15,706** | **39,176** |

***Table 2. Lead isotope ratios measured in 20 ingots having unknown geographical origin. Numbers from 1 to 13 are Roman ingots, while from 14 to 20 are Pre-Roman ingots (no mould-mark) dated 3$^{rd}$ century B.C. and cast in the typical shape using Pinna Nobilis (L.).***

TABLE 3
LEAD ISOTOPIC RATIOS MEASURED ON 24 ROMAN INGOTS
EPIGRAPHICALLY IDENTIFIED AS FROM CARTAGENA ORE DISTRICT.

Nos. 1 - 11 are reported from: Pinarelli *et al.* 1995. Ingot MVL617 bears an impressed stamp: PILIP. The Mal di Ventre ingots are kept at Oristano (Sardinia) while the Villasimius ingot is kept at Cagliari museum.
Nos. 12 - 23 are reported from Begemann, Schmitt-Strecker 1994, and ingot n.24 from Piccottini *et al.* 2002.

| *n* | Sample | Mould-mark | Site of discovery | Date of the wreck | Ri/j 206/207 | Ri/j 208/206 | Ri/j 206/204 |
|---|---|---|---|---|---|---|---|
| *1* | 1 | SOC. M.C. PONTILIENORVM.M.F | Mal di Ventre (Sardinia) Wreck | Mid 1st c. B.C. | 1,1932 | 2,0834 | 18,721 |
| *2* | 2 | L. CARVLI L.F.HISPALI MEN | Mal di Ventre | " | 1,1947 | 2,0827 | 18,722 |
| *3* | 3 | Q.APPI // *delphinus* // C.F *ancora* | Mal di Ventre | " | 1,1932 | 2,0863 | 18,731 |
| *4* | 4-MVL132 | L. (*or* M. ?) PLANI.L.F // *delphinus* // RVSSINI | Mal di Ventre | " | 1,1953 | 2,0825 | 18,735 |
| *5* | 5-MVL455 | M.C.PONTILIENORVM.M.F | Mal di Ventre | " | 1,1939 | 2,0845 | 18,729 |
| *6* | 6-MVL376 | CN.ATELLI.T.F.MENE | Mal di Ventre | " | 1,1941 | 2,0837 | 18,727 |
| *7* | 7-MVL617 | SOC.M.C.PONTILIENORVM.M.F | Mal di Ventre | " | 1,1942 | 2,0832 | 18,711 |
| *8* | 8-MVL456 | *caduceus* (?) // L.APPULEI L.L.PILON // *delphinus* | Mal di Ventre | " | 1,1940 | 2,0789 | 18,723 |
| *9* | 9-MVL377 | *caduceus* // M.PINARI.M.F // *delphinu*s | Mal di Ventre | " | 1,1941 | 2,0838 | 18,718 |
| *10* | 10-MVL618 | C.VTIVS.C.F // *delphinus* | Mal di Ventre | " | 1,1952 | 2,0827 | 18,732 |
| *11* | 11 | [.] PONTI[…..].M.F | Villasimius (Sardinia) no wreck | | 1,1944 | 2,0813 | 18,702 |
| *12* | MB52,1 | [……….] // [……….] // […….] | *Mahdia Wreck (Tunisia)* | c. 110-90 B.C. | - | 2,0831 | 18,748 |
| *13* | MB52,2 | M.PLANI L.F //*delphinus*// RVSSINI | *Mahdia* | " | 1,1952 | 2,0817 | 18,751 |
| *14* | MB52,3 | CN. ATELLI. T.F.MENE | *Mahdia* | " | 1,1946 | 2,0817 | 18,737 |
| *15* | MB52,4 | CN. ATELLI. T.F.MENE | *Mahdia* | " | 1,1942 | 2,0832 | 18,734 |
| *16* | MB52,5 | M.PLANIL.F //*delphinus*// RVSSINI | *Mahdia* | " | 1,1952 | 2,0818 | 18,734 |
| *17* | MB52,6 | M.PLANIL.F //*delphinus*// RVSSINI | *Mahdia* | " | 1,1950 | 2,0819 | 18,741 |
| *18* | MB52,7 | M.PLANIL.F //*delphinus*// RVSSINI | *Mahdia* | " | 1,1953 | 2,0814 | 18,769 |
| *19* | MB52,8 | M.PLANIL.F //delphinus// RVSSINI | *Mahdia* | " | 1,1950 | 2,0821 | 18,737 |
| *20* | MB52,9 | L.PLANIL.F. RVSSINI // *ancora* | *Mahdia* | " | 1,1945 | 2,0825 | 18,720 |
| *21* | MB52,10 | L.PLANIL.F. RVSSINI // *ancora* | *Mahdia* | " | 1,1946 | 2,0818 | 18,727 |
| *22* | MB52,11 | CN. ATELLI. T.F.MENE | *Mahdia* | | 1,1950 | 2,0818 | 18,734 |
| *23* | MB52,12 | M.PLANI L.F //*delphinus* // RVSSINI | *Mahdia* | | 1,1947 | 2,0813 | 18,723 |
| *24* | - | C.IVNI.L.F.PA[eti….] | *Magdalensberg* | | 1,1929 | 2,0857 | - |

| Lead ingots list | Places of discovery | Archaeological dating | Mould-mark | Evidence | | |
|---|---|---|---|---|---|---|
| | | | | Geografic proximity | Epigraphy | Lead isotopes |
| 0001, 0003 | *Cabrera 2 (Sp)* , *Îlot de Brescou (F)* | 3rd cent. BC | *no* | | | + |
| 1003 | *Mal di Ventre (Sard.)* | mid Ist cent. BC | Q.APPI // *delphinus* // C.F *ancora* | | | + |
| 1004 | « | « | *caduceus* (?) // L.APPULEI L.L.PILON // *delphinus* | | + | + |
| 1005 | *Escombreras 2 (F)*, S. Garraf Beach (Sp) | Ist cent. BC | *delphinus* // C.AQVINI // *ancora* | + | + | + |
| 1006 | *Bajo de Dentro (Sp)* | « | M. AQVINI.C.F. | + | + | + |
| 1008 | *Mahdia (T), Mal di Ventre (Sard)* | 100-50 BC« | CN.ATELLI.T.F.MENE | | + | + |
| 1009 | *Capo Testa B (Sard)* | c. 75-25 BC ( ?) | CN.ATELLI.CN L BVLIO | | + | |
| 1010 | Ischia (I) (no wreck) | | CN.ATELLI.CN.F.MISERINI | | + | + |
| 1012 | Harbour of Cartagena (Sp) | | L.AVRVNC L.[.]C[..]TA | + | + | |
| 1013 | Huelva mines (Sp)(no wreck) | | L. AVRVNC L.L.AT (*or* TA] | | + | + |
| 1015 | *Bajo de Dentro (Sp)* | Ist cent. BC | *lunula* // M.SEX.CALVI.M.F. // *lunula* | + | | + |
| 1016 | Ostra Vetere (I), Crestaig (Bal), *Îlot des Moines(Corsica) (F), Mal di Ventre (Sard)* | Ist cent. BC | L.CARVLI.L.F.HISPALI MEN | | | + |
| 1017 | *Madrague de Giens (F)* | 70-50 BC | L.CARVLI.L.F.HISPALLI MEN | | | + |
| 1018 | Roma (Tiber) (I) (no wreck) | | P.CORNEL.L.F AIM POLLION FORMIAN // GAL | | + | |
| 1019 | Harbour of Cartagena (Sp) | | M.DIRI MALCHIONIS *delphinus* | + | | |
| 1020 | *Escombreras 2 (Sp)* | Ist cent. BC | C.FIDVI.C.F // S.LVCRETI.S.F | + | | + |
| 1022 | Martigues (Bouches-du-Rhône) (F) (no wreck) | | A.P.FVRIEIS.C.P.L.L | | + | + |
| 1023 | Off Marseillan-Plage (F) (= Parker 16) | (late IInd-early Ist cent. BC | SOC. // L.GARGILI.T.F.ET.M.LAETILI.ML // *delphinus* | | + | + |
| 1078 | Magdalensberg (A) (no wreck) | | C.IVNI.L.F.PA[eti] | | + | + |
| 1030 *(lost)* | Harbour of Cartagena (Sp) | | LAETILI FERM *caduceus* | + | + | |
| 1031 | *Bajo de Dentro (Sp)* | Ist cent. BC | L.LVCRETI.L.F. C.M[….].F | + | + | |
| 1032 | Underwater, near Cartagena (Sp) | | [.].S.LVCRETI *hedera* // SOCIETATVS | + | + | + |
| 1033 | Monastery of Klingenthal (near Basel) (S) (no wreck) | | SOCIETAT // S.T.LVCRETI | | + | + |
| 1034 | *Bajo de Dentro (Sp)*, Savignano (I) | Ist cent. BC | *delphinus* // C. MESSI.L.F | + | + | + |
| 1035 | Harbour of Cartagena (Sp) | | P.NONAE.P.F.NVC | + | | + |
| 1036 *(lost)* | Harbour of Cartagena (Sp) | | *delphinus* // C.NONI.ASPRENATIS | + | | |
| 1037 *(lost)* | Riotinto mines (Sp) (no wreck) | | NOVA CARTHAGO | | + | |
| 1039 | *Mal di Ventre (Sard)* | Ist cent. BC | *caduceus* // M.PINARI.M.F // *delphinus* | | | + |
| 1040 | Cala Cartoe (Sard) | | L.PLAANI L.F // RVSSINI | | | |
| 1041 | Harbour of Cartagena (Sp), Cianciana (Sic) | | L.PLANI.LF. // *ancora* | + | | |
| 1042 | *Mahdia (T), Off Marseillan plage (F)* | Ist cent. BC | L.PLANI.L.F.RVSSINI // *ancora* | | | + |
| 1043 | *Bajo de Dentro (Sard), Mal di Ventre (Sard)*, Ripatransone (I). *Escombreras 2* (= 1043 bis) (Sp), harbour of Denia (Sp) | « | L.PLANI.L.F // *delphinus* // RVSSINI | + | | + |
| 1044 | *Mahdia (T), Mal di Ventre (Sard)*, Algalarens (Minorca) (Sp), Gavdos island (G) | « | M.PLANI L.F // *delphinus* // RVSSINI | | | + |
| 1046 | Baie de la Roquille (Agde) (F) | | C.PONTILIENI S.F. *delphinus* | | + | + |
| 1047 | Harbour of Cartagena (Sp), Off Palavas (F) | no date | C.PONTILIENI.M.F. *delphinus* | | + | |
| 1048 | Volubilis (M) | | C.PONTILIENI.M.F.FAB | | + | |
| 1049 | *Mal di Ventre (Sard)* | Ist cent. BC | M.C.PONTILIENORVM.M.F | | + | + |
| 1050 | *Mal di Ventre* (Sard), *Off Marseillan plage (F)* | Ist cent. BC | SOC.M.C.PONTILIENORVM.M.F | | + | + |
| 1051 | Villasimius (Sard)s | | [.]PONTI[…].M.F | | + | + |
| 1054 | Harbour of Cartagena (Sp) | | M.RAI.RVFI // caduceus // FER | + | + | + |
| 1055 | Caserio de Roches (La Unión) (Sp) | Ist cent. BC | M.PROSCIEIS.M.F.MAIC | + | | + |
| 1056 | *Nido del Cuervo (Sp)* | « | Q.SEI. P.F.MEN POSTVMI | | + | + |
| 1063 | Harbour of Cartagena (Sp) | | P.TVRVLLI LABEON // *delphinus* | + | + | |
| 1064 | Harbour of Cartagena (Sp) | | P.TVRVLLI.M.F. // MAI *delphinus* | + | + | + |
| 1071 | Harbour of Cartagena (Sp), Harbour of Cherchel (Ar) | | Q VARI HIBERI | + | + | + |
| 1072 | *Madrague de Giens* (F), Punta Falcone (Sard), *Mal di Ventre(Sard)*, fiume Stella (I) | 70-50 BC | C.VTIVS.C.F // *delphinus* | | | + |
| 1073 | *Bajo de Dentro(Sp)*, Ventotene (I) | 1st cent. BC | *delphinus* // C.VTIVS.C.F. // *caduceus* | | | + |
| 1074 | Harbour of Cartagena (Sp), *Capo Testa B (Sard)* | c. 75-25 BC ? | C.VTI.C.F.MENEN | + | | |
| 1503 | *Baie de l'Amitié* (Marseillan) (F) | 50-150 AD | (no mould-mark) | | | + |
| 1507 | Riotinto mines (Sp) no wreck) | | (no mould-mark) | | | + |
| 1702 | *Danger d'Algajola (Corsica) (F)* | c. 150-100 BC | […..] // […..] | | | + |
| 1703 | *Ilot des Moines (Corsica) (F)* | no date | […..] // […..] | | | + |
| 1704 | Golfe de Ventilegne (Corsica) (F) | no date | [….] // [……….] // [….] | | | + |
| 1705 | Off Grande Sanguinaire (Corsica) (F) | no date | [….] // [……….] // [….] | | | + |
| 1706 | *Madrague de Giens (F)* | 70-50 BC | […..]I.L.F.[…..] | | | + |
| 5001 | *Comacchio (Ferrara) (I)* | Late Ist cent. BC | (stamps impressed) | | + | + |

***Table 4. SUMMARY TABLE. Lead ingots coming from Cartagena silver-lead mines according to: the geography of the discovery, the epigraphy and the lead isotope ratios measurement.***
***In Italic : the identified wrecks, the date of which is given by the composition of the cargoes (see Parker 1992). The dolphin marked on the ingots number 1043 is upside down (head on the right) while on the ingots number 1043 bis it is upside down again, but with the head on the left.***
***A : Austria ; Ar : Argelia ; Bal : Balearic Islands ; F : France ; G : Greece ; I : Italy ; M : Morocco ; S : Switzerland ; Sard : Sardinia ; Sic : Sicily ; T : Tunisia.***

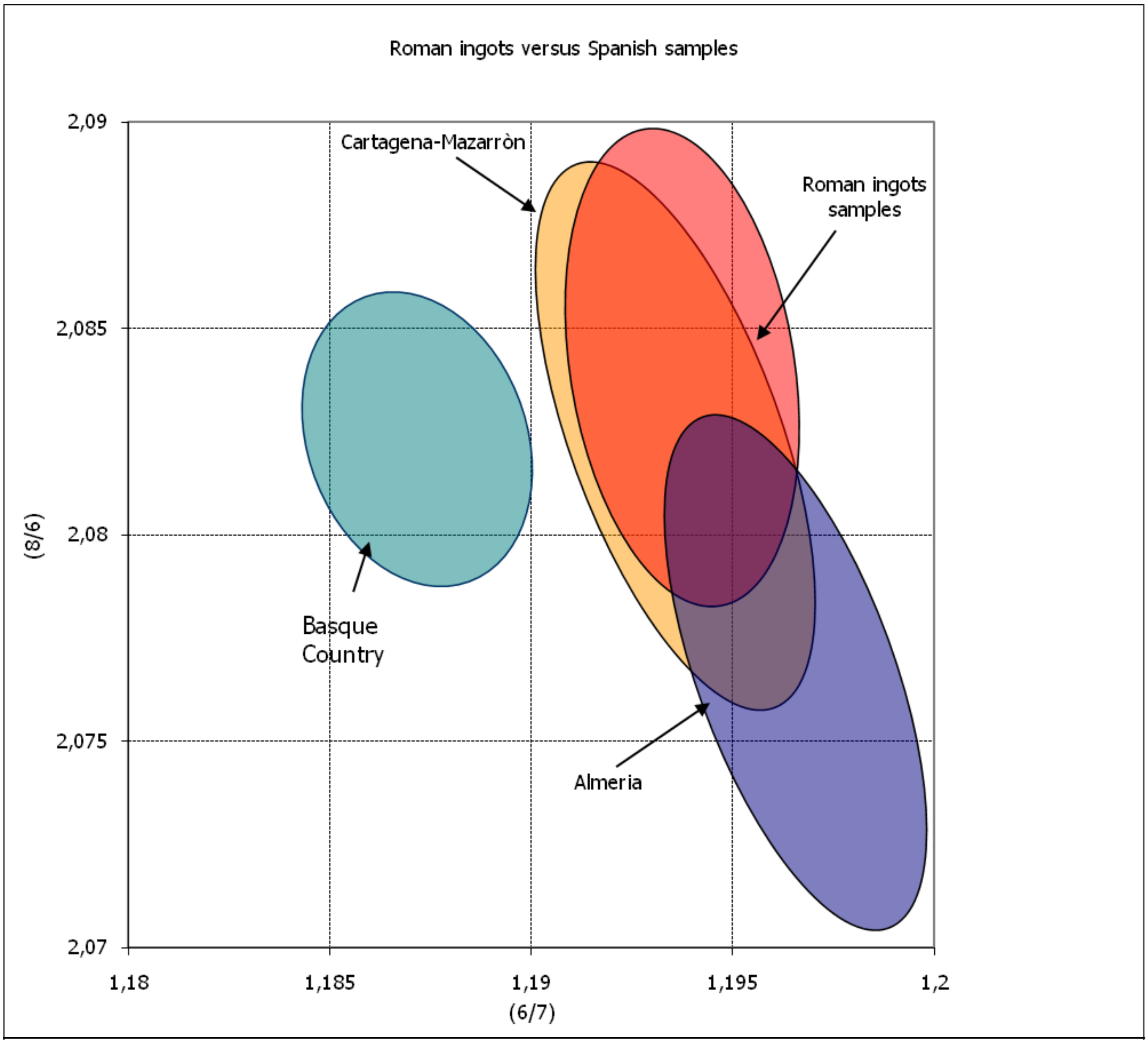

Graph 5. Spanish cluster areas. The cluster defined by 66 Roman lead ingots having their origin in Cartagena is very close to the cluster formed by the lead mineral samples collected in the Cartagena ore district. The small differences between these two clusters are probably due to the mass spectrometry measurements which are not homogeneous (some old measurements can be affected by higher uncertainties).

Graph 3 clearly highlights why these ingots are to be included in the Cartagena cluster. The group in this list of Roman and pre-Roman ingots of unknown geographical origin can be examined:

*Pre-Roman Ingots*

The 7 pre-Roman ingots were cast in their characteristic shape using *Pinna Nobilis (L.)* shells (fig.4). They weigh *c.*40-42 kg. Four of them belong to a cargo discovered in the wreck *Cabrera 2* (Balearic Islands) dating to *c.* 220-225 B.C.; three are now kept in the monastery of Lluc, Mallorca. Another wreck (Ïlot de Brescou near Agde) of the 3rd c. B.C. contained a cargo of lead ingots of this type; one of them is preserved in the Musée du Vieux-Biterrois (Béziers) and two others probably in the Musée de L'Éphèbe, at Cap d'Agde. The lead isotope ratios (Tab.2) show a good compatibility with the Cartagena values. This confirms the fact that this area was partially exploited before the Roman settlement, an opinion expressed nearly 20 years ago (Domergue 1990, 167-68), when lead isotope analysis was not a common practice.

*Roman Ingots*

i) Samples 99.24 and 02.06

The first ingot was discovered at *La Madrague de Giens* (Var) dating to 70-50 B.C. and is now kept in DRASSM (Fort Saint-Jean, Marseilles), while the second is preserved in the Museum of Rio Tinto. The measurement of the lead isotopic ratios shows that both of these ingots originated in the Cartagena ore district.

ii) Samples 00.09, 00.10 and 00.11

All three samples belong to the same wreck discovered at *Baie de l'Amitié* (Marseillan, Hérault) dating to A.D. 50-150 with 98 ingots on board; they are now kept in the Musée de l'Ephèbe, Cap d'Agde. Measurement of the lead isotope ratios identifies these samples too as belonging to the Cartagena cluster.

iii) Samples from Corsica

Samples 98.29, 98.30 and 98.31 belong to a wreck discovered at *Domain d'Algajola* (Calvi, Corsica) dating to 150-100 B.C., where a submarine plateau extends from 1m to 40m depth. Samples 98.10, 98.32 and 98.33 come from the island of *Îlots des Moines (Bonifacio, Corsica)* (no dated). The remaining samples, 98.22 and 98.23, were both recovered off the coast of Ajaccio (no dating). All these samples also belong to the Cartagena cluster.

## Summary and conclusions

From our database we drew an initial representative cluster of 43 samples of ingots epigraphically proved to be Roman. Both in order to confirm their validity and to add other elements that would lead to a clear definition of the Cartagena cluster, we also considered 11 other samples of ingots geographically recognized as coming from Cartagena, insofar as they are stamped with marks traceable to Roman families active in the mines there. The isotopic measurements were performed by Pinarelli *et al.* (1995). These 11 ingots formed part of the cargo recovered from the wreck of a ship that sank close to the island of *Mal di Ventre*, Sardinia in the mid-1$^{st}$ c. B.C. (Tab.3). Subsequently we added the isotopic measurements of 12 other ingots originating from the Cartagena mines and stamped with the names of well-known Roman manufacturing families. These data, published by Begemann and Schmitt-Strecker (1994), refer to the Mahdia (Tunisia) wreck dated to *c.*110-90 B.C. (Table 3). Graph 4 shows the new cluster, defined by 66 identified ingots.
In relation to the cluster area referred to above (Graph 5), and from a statistical point of view, we can now establish a very representative value for the lead isotopic ratios (the "*isotopic signature*") characteristic of the Cartagena ore district :

$$^{206/207}Pb = 1.1947 \pm 0.0008 \qquad ^{208/206}Pb = 2.0826 \pm 0.0017$$

From an analytical perspective, this work defines the graphical cluster and the numerical isotopic ratios which make it possible to attribute the provenance of hitherto-unidentified samples of lead to the Cartagena ore district.

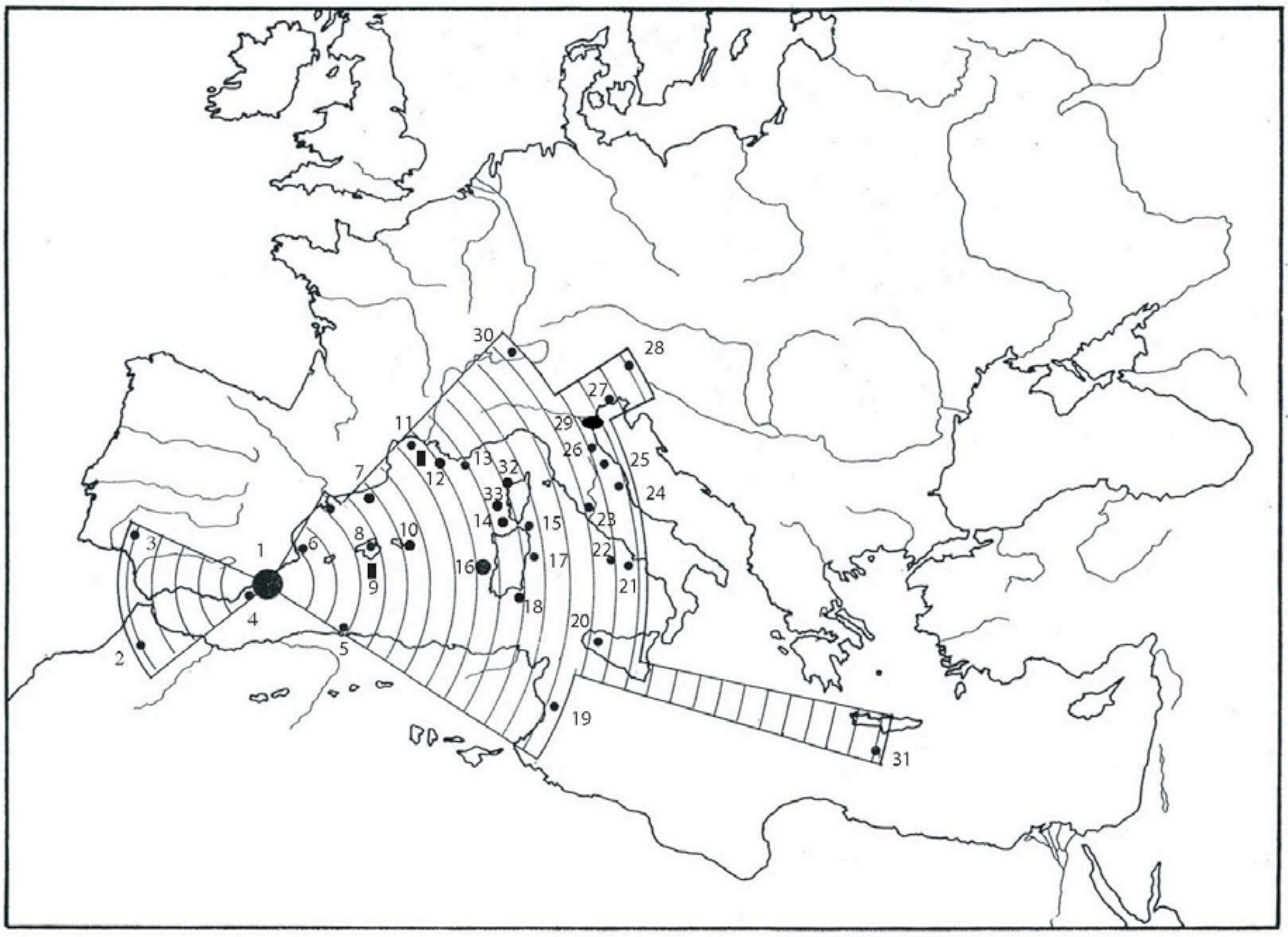

Fig.5 Diffusion of lead ingots from Cartagena (wrecks and findings on land).
Bar: Pre-Roman era (5th-3rd c. B.C.); Point: end of 2nd-1st c. B.C.; Ellipse: age of Augustus.

1. Cartagena (harbour, Escombreras, vicinity of La Unión, Cabo de Palos).
2. Volubilis (Morocco).
3. Huelva (Riotinto).
4. Aguilas (Nido del Cuervo) (Almería).
5. Cherchell (Algeria).
6. Denia (harbour) (Alicante).
7. Sierra de Garraf beach (Tarragona).
8. Mallorca (Crestaig).
9. Cabrera Island.
10. Minorca Island.
11. Marseillan (Riches-Dunes, Baie de l'Amitié) – Agde (Baie de la Roquille, îlot de Brescou) – Frontignan – Palavas.
12. Martigues.
13. Giens (La Madrague).
14. Îlots des Moines and golfe de Ventilegne (Corse du Sud).
15. Bouches-de-Bonifacio (Capo Testa, Punta Falcone, Sardinia).
16. Mal di Ventre (Sardinia).
17. Cala Cartoe (Dorgali, Sardinia).
18. Villasimius (Sardinia).
19. Mahdia (Tunisia).
20. Cianciana (Sicily).
21. Ischia (Italy).
22. Ventotene Island (Italy).
23. Rome.
24. Ripatransone (Ascoli-Piceno, Italy).
25. Ostra Vetere (Ancona, Italy).
26. Savignano (Italy).
27. Fiume Stella (Italy).
28. Magdalensberg (Austria).
29. Comacchio (Ferrara, Italy).
30. Bâle.
31. Gavdos Island (Greece).
32. Danger d'Algajola (Haute Corse, F).
33. Îles Sanguinaires (Corse).

As our database contains many other isotopic results from different lead samples originating in the Iberian peninsula, we intend in the near future to generate clusters in connection with other major mining districts.

Our findings are highly positive, first as regards the methods used. There is no divergence on the one hand between the isotope data provided by various laboratories (three in the case of the Planii or Pontilieni ingots, for example: Tables 1 and 3), and on the other more generally between the isotope data and the results emerging from archaeology and epigraphy. The epigraphic method, sometimes aided by geographic observations (findspot near the mines, for example), often yields good results, at least when the ingots bear legible stamps, with the names of the producers, and if there is a good data bank of epigraphic references. This is the case with *Carthago Noua*, but matter are not always so straightforward: for example, as regards ingots from the Sierra Morena mines, the epigraphic line of reasoning may be inadequate, although those ingots too carry the producers' stamps, as the mines in that region were not so clearly associated with an urban centre like *Carthago Noua*.

Yet, as regards ingots where the *nomen* of the producer cannot be tied to *Carthago Noua*, the epigraphic argument has no validity. That is the case with the ingots stamped with the names of one Gaius Appius or one Marcus Pinarius (Tables 3-4). To be sure, the uniformity of the cargo can be invoked (the cargo of the Mal di Ventre wreck of the mid-1st c. B.C. is a good example), but this may be risky, as cargoes of metal may have been made up of ingots of different origins. Under these conditions the isotopic method will carry more weight. For a variety of reasons, however, the isotopic argument may be weakened. This is the case with the ingot of the Magdalensberg produced by a certain Gaius Iunius, but in this case the epigraphic argument may come to the assistance of the isotopic, and may here be the deciding factor (Domergue and Piccottini 2004). Finally, when ingots are anepigraphic, as with pre-Roman ingots (Table 2), the isotope method is the only feasible one. The same holds true if the mould-marks are lacking or have been erased (Table 2). In short, the two methods of investigation gain strength from a comparison between them.

The combined approach yelds a clearer picture of the trade in lead from the *Carthago Noua* mines (Table 4 and fig. 5). It emerges that the period of production and trade of this lead throughout the Mediterranean world was longer than was previously thought, particularly in the earlier period. Now we may be certain that it was exported as far back as the 3rd c. B.C. This dating will undoubtedly be narrowed down further in the future, but at least it shows that lead from *Carthago Noua* was transported along the sea routes well before the Roman period. It is also true that the peak of lead production from *Carthago Noua* occurred at the end of the 2nd and especially in the 1st century B.C., as illustrated by the magnificent ingots from the *Mal di Ventre* wreck of the mid-1st c. B.C. or those from the *Escombreras 2* wreck at Cartagena of the 1st c. B.C. (fig. 2). Until now, the ingots from *Comacchio* wreck of the late 1st c. B.C. (fig. 3) appeared to mark the end of this period, seeming to provide a *terminus ante quem* in the Augustan age, in *c.* 15-10 B.C. (Domergue *et al.* forthcoming). However, measurements of the lead isotope ratios of samples from three lead ingots from *Baie de l'Amitié* wreck at Marseillan (Table 2) identifies them as belonging to the Cartagena cluster. That poses a problem because of the dating of the wreck (A.D. 50.150), making it the only group of ingots from *Cartago Noua* mines later than the 1st c. B.C.[4]

During the 2nd and 1st c. B.C., the lead producers from *Carthago Noua* established a regular trading network in the W Mediterranean, as the results of our multidisciplinary investigation show (fig.5). Our starting point was not promising (*Cabrera 2,* from the Balearics and *Îlot de Brescou* at Agde, both of the 3rd c. B.C.), but those wrecks were already located on two of the great sea routes of the end of the Late Repubblic: the Spain-Italy link via Balearics and the coastal route. With progress in research, and with input from epigraphy as well as analyses (e.g., of ingots without trademarks found off the coasts of Corsica), these routes and others are becoming clearer. The ingots from Basle and Magdalensberg testify to the entry of Spanish lead into continental Europe. However, the ingots from the Gavdos wreck (1st c. B.C.) are probably nothing more than a coincidence: they were very likely part of the ship's stores, used for the repairs which were from time to time necessary on a ship on the high seas trading with the E Mediterranean, rather than the remains of a cargo of metal destined for that far-off region.

[4] Here we simply indicate the problem, to be treated elsewhere